# Ferrimagnetos de ising con dilución*

*Nuri Hurtado*[1]** *y Gloria María Buendía*[2]

[1]*Centro de Física Teórica y Computacional. Facultad de Ciencias, Universidad Central de Venezuela (UCV). Caracas, 1041A, Venezuela.* [2]*Departamento de Física, Universidad Simón Bolívar (USB). Caracas, 1080, Venezuela*

## Resumen

Hemos realizado un estudio numérico de un modelo mixto de Ising en 3 dimensiones diluido con diferentes concentraciones de impurezas no magnéticas, formando un desorden tipo *quenched* en el sistema. Nuestros resultados preliminares indican que tanto la temperatura de compensación del sistema, como la temperatura a la cual el sistema pasa de una fase ordenada a una fase desordenada, decaen en forma lineal con la concentración de impurezas, para un porcentaje de impurezas inferior al 30%, por encima de este valor la temperatura de compensación desaparece.

**Palabras clave:** Ising; ferrimagnetismo; monte carlo; temperatura de compensación; quenched.

# Diluted ising ferrimagnets

## Abstract

We perform numerical studies of a three dimensional Ising mixed model at different concentrations of quenched non magnetic impurities. Our preliminary results indicate that both, the compensation temperature and the temperature that signal the transition from an ordered to a disordered phase, decrease linearly with the concentration of impurities, for concentrations below 30%. For percentages of impurities above 30% the compensation temperature disappears.

## Introducción

Los modelos mixtos de Ising han sido utilizados con éxito para simular varios aspectos del comportamiento magnético de materiales ferrimagnéticos, entre ellos la existencia de temperaturas de compensación $T_{comp}$ (1, 2). *A la* $T_{comp}$ la magnetización total del sistema se anula por que las magnetizaciones de las subredes se cancelan, a diferencia del punto crítico $T_c$ en el que tanto la magnetización total como las magnetizaciones de las subredes se hacen cero. La existencia de la $T_{comp}$ tiene importantes aplicaciones tecnológicas principalmente en el área de grabado termomagnético de información (3).

Trabajos previos indican que la $T_{comp}$ es muy sensible a las interacciones entre moléculas vecinas, así como también al campo cristalino y a los campos magnéticos externos (1-4, 5). En general como el comportamiento de un sistema puro es muy diferente al de un sistema con impurezas (6), es de esperarse que la presencia de éstas afecte la temperatura de compensación. Conocer esta dependencia puede tener interesantes aplicaciones prácticas, ya que variando el porcentaje y el tipo de impurezas se podría variar el rango de la temperatura de compensación.

## Modelo y Métodos

En una red cúbica de tamaño $L^3$, con $L= 24$, se colocan en forma alternada espines de tipo $\sigma = \pm 1/2$ y $S = 0, \pm 1$. El Hamiltoniano que describe este sistema viene dado por la expresión,

$$H = -J_1 \sum_{<nn>} (\sigma_i \varepsilon_i)(S_j \varepsilon_j) - J_2 \sum_{<nnn>} (\sigma_i \varepsilon_i)(\sigma_k \varepsilon_k) \quad [1]$$

donde $J_1$ y $J_2$ son las interacciones a primeros y segundos vecinos, $D$ y $H$ representan el campo cristalino y el campo externo respectivamente, todos en unidades de energía. Los valores de los parámetros: $J_1 = -0,5$, $J_2 = 1,00$, $D = 1,00$ y $H = 1,00$ se mantuvieron fijos para todas las muestras. Los factores $\varepsilon_i$, introducen la dilución, ya que $\varepsilon_i = 0\ (1)$ equivale a la ausencia (presencia) de un espín en el sitio i de la red. El valor de $\varepsilon_i$ se elige aleatoriamente. Definimos $\lambda$ como el número promedio de impurezas "quenched" en la red por sitio, este valor está asociado al número de ceros "fijos" en la red debidos a la ausencia de un espín, no debe confundirse con los espines S que eventualmente pueden tomar el valor cero.

En este trabajo preliminar se utiliza una red relativamente pequeña, por lo que nuestros resultados presentarán evidentes efectos de tamaño finito, sin embargo la temperatura de compensación, al no ser un valor crítico, es menos afectada por estos efectos.

Se usa el método de Monte Carlo con $4x10^3$ pasos para calentar el sistema y $4x10^4$ pasos para el cálculo de promedios configuracionales. Se hace un barrido secuencial de la red para generar las configuraciones, utilizando el algoritmo de baño térmico.

Para cada valor de $\lambda$, se repite el procedimiento para 15 distribuciones distintas de impurezas. Cada una de esas distribuciones permanece fija respecto a variaciones de temperatura y tiempo, lo que en la literatura se conoce como desorden tipo "*quenched*". Se calculan: la magnetización total del sistema,

$$m = [\langle M \rangle] = [\langle M_\sigma \rangle + \langle M_S \rangle] \quad [2]$$

las magnetizaciones de las subredes,

$$m_\sigma = [\langle M_\sigma \rangle]$$
$$m_S = [\langle M_S \rangle] \quad [3]$$

donde

$$\langle M_\sigma \rangle = \sum_i \sigma_i \varepsilon_i$$
$$\langle M_S \rangle = \sum_j S_j \varepsilon_j \quad [4]$$

y la susceptibilidad magnética,

$$x = [\beta(\langle M^2 \rangle - \langle M \rangle^2)] \quad [5]$$

para diferentes valores del número promedio de impurezas por sitio, $\lambda$, donde $[\ ]$ indica el promedio sobre distribuciones de desorden, $\langle\ \rangle$ el promedio térmico y $\beta = 1/(K_B T)$, la constante de Boltzman $K_B$ se tomará como 1 a lo largo de este trabajo.

La $T_{comp}$ se obtiene cuando las magnetizaciones de las subredes son iguales en módulo y de signos opuestos

$$|m_\sigma(T_{comp})| = |m_S(T_{comp})|$$
$$signo(m_\sigma(T_{comp})) = -signo(m_S(T_{comp})) \quad [6]$$

es decir, que para esta temperatura la magnetización total del sistema se hace cero.

## Resultados y Discusión

En ausencia de campo externo la temperatura a la cual la susceptibilidad diverge indica en general una transición en el sistema de una fase ordenada a otra desordenada. El sistema estudiado tiene un campo externo, por lo que el máximo de la susceptibilidad no necesariamente indica un cambio de fase. Nuestros resultados indican que la susceptibilidad de nuestro modelo posee dos máximos, el de menor altura ocurre a una temperatura $T_m$ superior a la temperatura a la que ocurre el pico de mayor altura. A $T_m$ se observa que el sistema va de una fase ordenada a otra desordenada (4). Para temperaturas superiores a $T_m$ la magnetización total del sistema y las magnetizaciones de las subredes son siempre cero.

En la Figura 1 mostramos la altura del segundo máximo de la susceptibilidad vs. el número promedio de impurezas por sitio $\lambda$. La línea indica un ajuste de tipo $\chi_{max} \propto \lambda^{-1/2}$, que fue el mejor ajuste encontrado para estos datos. Estudios numéricos previos de modelos de Ising con impurezas en 2 y 3 dimensiones (7, 8) muestran, al igual que en nuestro modelo, una disminución en la altura y la posición de los máximos de susceptibilidad con el aumento de la concentración de impurezas.

En la Figura 2 mostramos la dependencia de $T_m$ (mínima temperatura a la cual todas las magnetizaciones van a cero), y de la temperatura de compensación $T_{comp}$, respecto el número promedio de impurezas por sitio, $\lambda$. Un ajuste lineal de $T_m$ nos da la relación $T_m = (-3,0 \pm 0,1)\lambda + (2,83 \pm 0,04)$. Encontramos que el valor obtenido como punto de corte es del orden del valor numérico que se obtiene en otros trabajos del mismo modelo pero sin impurezas (4), $T^*_m = (2,65 \pm 0,05)$. Encontramos que existe una concentración de impurezas del orden del 30% para la cual ambas temperaturas coinciden. Por encima de este valor $T_{comp}$ desaparece. Un ajuste lineal nos dice que $T_{comp} = (-1,45 \pm 0,08)\lambda + (2,25 \pm 0,01)$, que coincide dentro de los errores estadísticos con valores previos para el caso puro (4), $T^*_{comp} = (2,20 \pm 0,05)$. Los valores para el sistema puro fueron obtenidos para redes de mayor tamaño, lo cual parece indicar que pese a que el ta-maño de la red en este trabajo es pequeño, los resultados son relativamente confiables dentro de un margen de error.

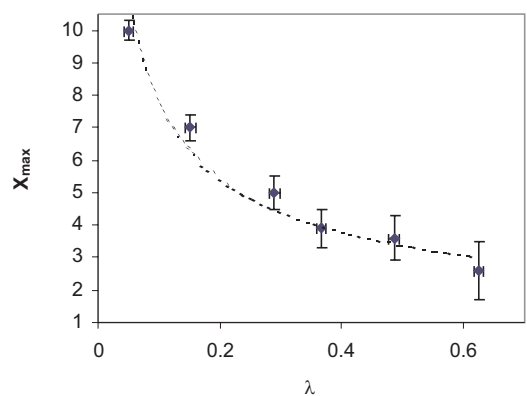

Figura 1. Altura de los máximos de susceptibilidad en función del número promedio de impurezas en el sistema. La línea representa un ajuste de tipo $\chi_{max} \propto \lambda^{-1/2}$.

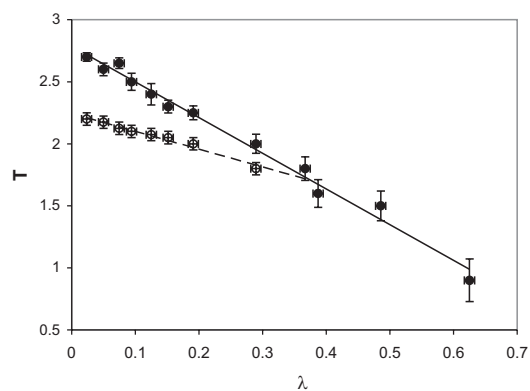

Figura 2. Temperatura de compensación (círculos vacíos), $T_{comp}$, y temperatura mínima a la cual todas las magnetizaciones se hacen cero (círculos llenos), $T_m$, en función del número promedio de impurezas del sistema. Las líneas representan los ajustes de tipo lineal.

## Conclusiones

En este trabajo preliminar, hemos utilizado el método de Monte Carlo para determinar la temperatura de compensación, $T_{comp}$, y el comportamiento de la altura de los máximos de la susceptibilidad magnética y la temperatura a la cual ocurren estos máximos, $T_m$, en un modelo mixto de Ising tridimensional, en presencia de diferentes concentraciones de impurezas, quenched, no magnéticas distribuidas aleatoriamente sobre la red.

Encontramos que la altura del segundo máximo de susceptibilidad (el que corresponde al cambio de una fase desordenada a una ordenada) parece decaer con el inverso de la raíz cuadrada del número promedio de impurezas por sitio, $\lambda$, y que la temperatura a la cual se obtienen estos máximos de susceptibilidad disminuye de forma lineal con $\lambda$. Estos resultados, aunque preliminares debido al tamaño del sistema, son consistentes con los obtenidos por otros autores (7, 8). Nuestros resultados indican que, la temperatura de compensación, $T_{comp}$, también disminuye linealmente con el aumento de la concentración de impurezas, este resultado es novedoso y puede tener interesantes aplicaciones. Encontramos que para un porcentaje de impurezas, de aproximadamente *30%*, ambas temperaturas se hacen comparables. Para porcentajes mayores de impurezas la temperatura de compensación desaparece.

## Referencias Bibliográficas